\documentclass[aps,floatfix,twocolumn,final,prl,showpacs]{revtex4}

\usepackage{epsfig}
\usepackage{graphicx}
\usepackage[ansinew]{inputenc}

\begin{document}

% \preprint{APS/123-QED}
\preprint{APS/123-QED}
\title{Crystal structure and physical properties of EuPtIn$_{4}$ intermetallic antiferromagnet}% Force line breaks with \\

\author{P. F. S. Rosa,$^{1,2}$ C. B. R. de Jesus$^{1}$, Z. Fisk$^{2}$ and P. G. Pagliuso$^{1}$}

\affiliation{$^{1}$Instituto de F\'isica \lq\lq Gleb Wataghin\rq\rq,
UNICAMP, Campinas-SP, 13083-859, Brazil.\\
$^{2}$University of California, Irvine, California 92697-4574,
USA}

%\author{Charlie Author}
%  %\homepage{http://www.Second.institution.edu/~Charlie.Author}
% %\affiliation{
% Second institution and/or address\\
% This line break forced% with \\
% }%

\date{\today}% It is always \today, today,
             %  but any date may be explicitly specified

\begin{abstract}
We report the synthesis of EuPtIn$_{4}$ single
crystalline platelets by the In-flux technique. This compound crystallizes in the orthorhombic Cmcm structure with lattice parameters $a=4.542(1)$ \AA, $b=16.955(2)$ \AA$\,$ and $c=7.389(1)$ \AA. Measurements of magnetic susceptibility, heat capacity, electrical resistivity, and electron spin resonance (ESR) reveal that  EuPtIn$_{4}$ is a metallic Curie-Weiss paramagnet at high temperatures and presents antiferromagnetic
(AFM) ordering below $T_{N}=13.3$ K. In addition, we observe a successive anomaly at $T^{*} = 12.6$ K and a spin-flop transition at $H_{c} \sim 2.5$ T applied along the $ac$-plane. In the paramagnetic state, a
single Eu$^{2+}$ Dysonian ESR line with a Korringa relaxation rate of $b = 4.1(2)$ Oe/K is observed. Interestingly, even at high temperatures, both ESR linewidth and electrical resistivity reveal a similar anisotropy. We discuss a possible common microscopic origin for the observed anisotropy in these physical quantities likely associated with an anisotropic magnetic interaction between Eu$^{2+}$ 4$f$ electrons mediated by conduction electrons.
\end{abstract}

\maketitle

\section{\label{sec:intro}Introduction}

Low-dimensional rare-earth based intermetallic compounds exhibit a variety of interesting phenomena including 
Ruderman-Kittel-Kasuya-Yoshida (RKKY) magnetic interaction, heavy fermion (HF)
behavior, unconventional superconductivity, crystalline electrical field (CEF) and Fermi surface (FS) effects \cite{Zach,Szytula,Review}. In order to systematically explore the interplay between such versatile physical properties 
in structurally related series, it is highly desirable to separate the role of the each interaction in determining the behavior of the system.
For instance, the study of 
isostructural magnetic analogs have been often employed to elucidate the role of RKKY interactions and CEF effects in the evolution of the magnetic properties in $R_{m}M_{n}$In$_{3m+2n}$ 
($R=$ rare-earth; $M=$ Rh, Ir; $m=0, 1$; $m = 1, 2$) series \cite{Pagliuso,Granado}. In particular, Gd$^{3+}$- and Eu$^{2+}$-based members are usually taken as reference compounds due to their $S$-state ($S=7/2$, $L=0$)
ground state. As such, CEF effects
are higher order effects and their magnetic properties purely
reflect the details of RKKY interaction and FS
effects.

Among the Indium-rich compounds, the
series $RM$In$_{4}$ (114 system; $R$ = e.g. Ca, Eu, Yb, Ce; $M$ = e.g. Ni, Pd, Au) adopt the orthorhombic YNiAl$_{4}$-type structure which contains complex [PtIn$_{4}$] polyanionic networks with europium atoms filling distorted hexagonal channels (Fig. 1a) \cite{Po,EuAuIn4}. The clear elongation of the $b$-axis indicates the possibility of a $2$D Brillouin zone with cylindrical Fermi surfaces along the $b$ direction. Although the member CeNiIn$_{4}$ has been reported to display most likely a three-dimensional electronic state \cite{Onuki}, the promising features of this series of compounds have not been extensively explored yet, particularly for single crystalline samples. In order to test the above hypothesis in new members of this series, we here report the synthesis and physical properties of EuPtIn$_{4}$ single crystals. We have carried out electrical resistivity, magnetic susceptibility, specific heat and electron spin resonance (ESR) measurements. 
 The field-dependent magnetic susceptibility shows an AFM ordering at $T_{N}=13.3$ K  followed by a successive transition at $T^{*} = 12.8$ K. Both electrical resistivity and ESR linewidth are found to be  anisotropic even at high temperatures, suggesting the presence of an anisotropic magnetic interaction between the Eu$^{2+}$ 4$f$ electrons mediated by conduction electrons ($ce$).

\section{\label{sec:exp}Experimental Details}

Single crystalline samples of EuPtIn$_{4}$ were grown using
flux technique with starting composition Eu:Pt:In=1:1:25. The mixture was placed
in an alumina crucible and sealed in a quartz tube under vaccum. The sealed tube
was heated up to $1100\,^{\circ}\mathrm{C}$ for $2\,$h, cooled down
to $800\,^{\circ}\mathrm{C}$ at $20\,^{\circ}\mathrm{C}$/h and then cooled down to $350\,^{\circ}\mathrm{C}$ at $10\,^{\circ}\mathrm{C}$/h. The flux
was then removed by centrifugation and the obtained shiny platelet crystals are stable in air and have typical dimensions of
$1$ mm x $1$ mm x $0.1$ mm, as shown in Fig. 1b. Phase purity was checked by X-ray powder diffraction using a Rigaku diffractometer (Cu-K$\alpha$ radiation). Figure 1c shows the pattern of EuPtIn$_{4}$, which could be completely fitted with a single phase. Rietveld refinements of EuPtIn$_{4}$  ($R_{wp} = 13.6\%$) yields lattice parameters $a=4.542(1)$ \AA, $b=16.955(2)$ \AA$\,$ and $c=7.389(1)$ \AA. Specific heat measurements were
performed in a Quantum Design PPMS small-mass calorimeter that
employs a quasiadiabatic thermal relaxation technique. The electrical resistivity was measured using a
standard four-probe method also in the Quantum Design PPMS. The
magnetization was measured using a VSM superconducting quantum
interference device (SQUID) magnetometer  (Quantum Design).
ESR measurements were performed in a BRUKER spectrometer equipped
with a continuous He gas-flow cryostat. X-Band ($\nu$ $\sim 9$ GHz)
frequency was used in the
temperature region $4.2$ K $< T < 300$ K.

\section{\label{sec:results}Results and Discussion}

\begin{figure}[!ht]
\includegraphics[width=0.45\textwidth,keepaspectratio]{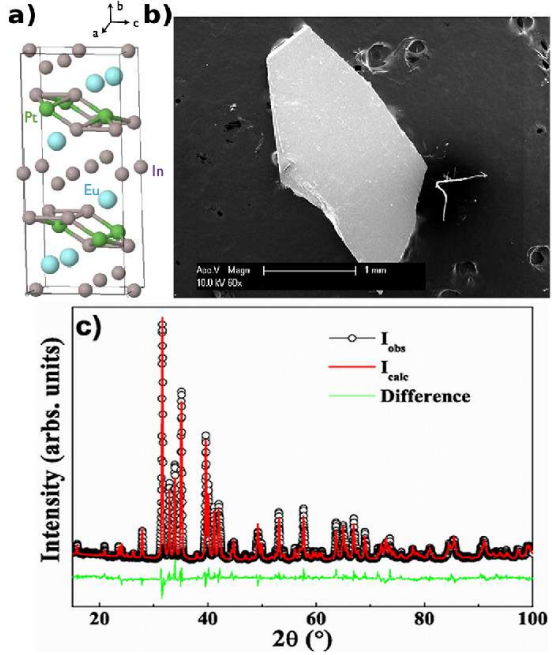}
\caption{a) Orthorhombic crystal structure of EuPtIn$_{4}$  (space group Cmcm). b) Scanning electron microscope (FE-SEM) image of as-grown EuPtIn$_{4}$ single crystal. c) X-ray powder pattern and Rietveld fit ($R_{wp} = 13.6\%$) of EuPtIn$_{4}$ at $300$ K.}
\end{figure}

The macroscopic physical properties of our EuPtIn$_{4}$ single crystals are presented in Fig. 2.
Panel a) displays the zero-field in-plane electrical resistivity, $\rho_{ac}$ ($T$), and aking the b-axis, $\rho_{b}$ ($T$), as a function of temperature. A weakly anisotropic metallic behavior is observed in the paramagnetic regime followed by a clear peak at $T_{N}$ = 13.3 K.  Residual resistivity ($\rho_{0}$) and residual resistivity ratio (RRR) values of $\rho$($T$) are $0.1-0.5$ $\mu\Omega$.cm and $\sim 70$, respectively, indicating good cristallinity of our samples. However, the magnetoresistance
 (MR $= \Delta\rho/\rho = \rho(H) - \rho(H=0)/\rho(H=0)$) at $T = 2$ K is linear with magnetic field and no quantum oscillations have been found (inset of Fig. 2a) up to $14$ T.  

\begin{figure}[!ht]
\includegraphics[width=0.4\textwidth]{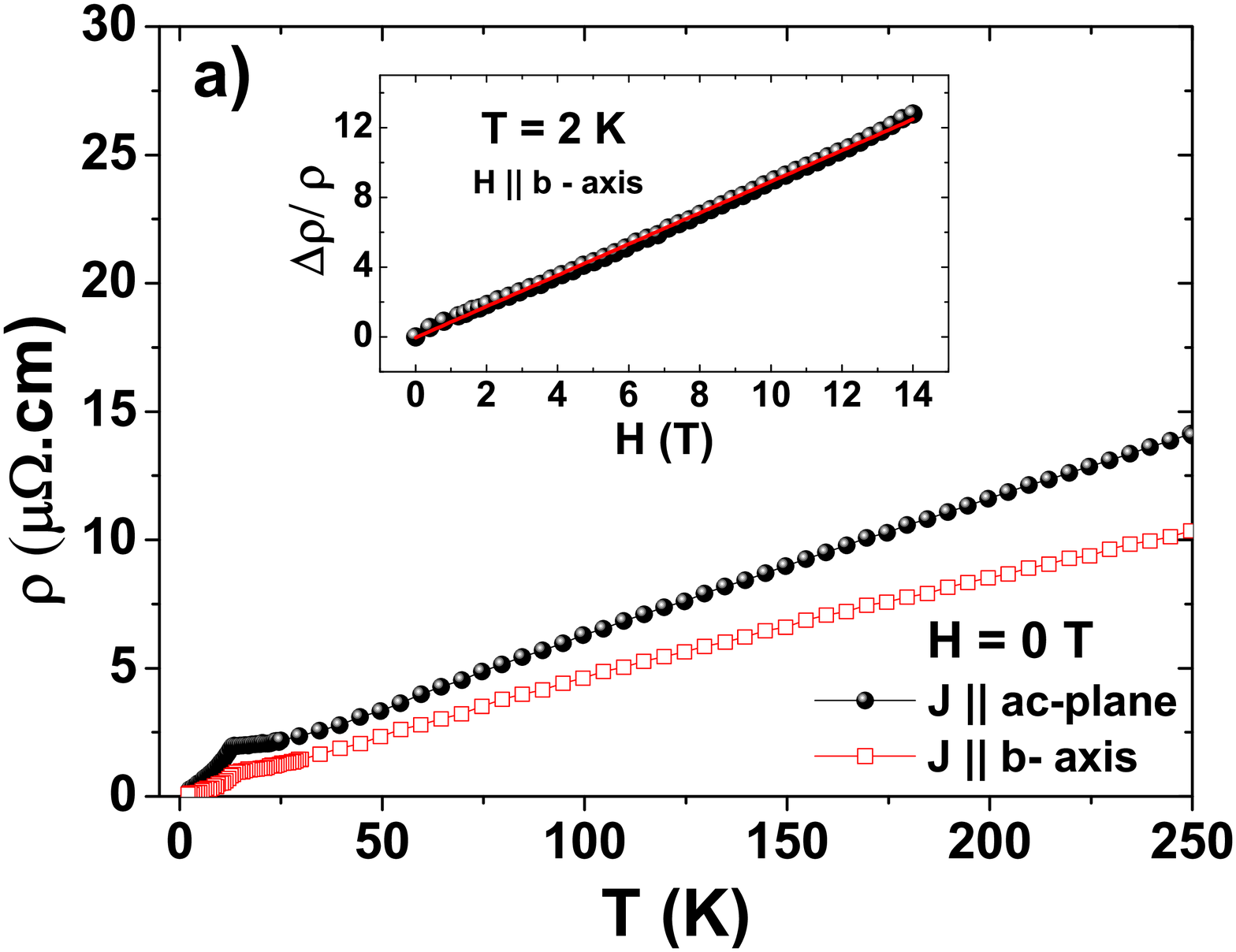}
\includegraphics[width=0.4\textwidth]{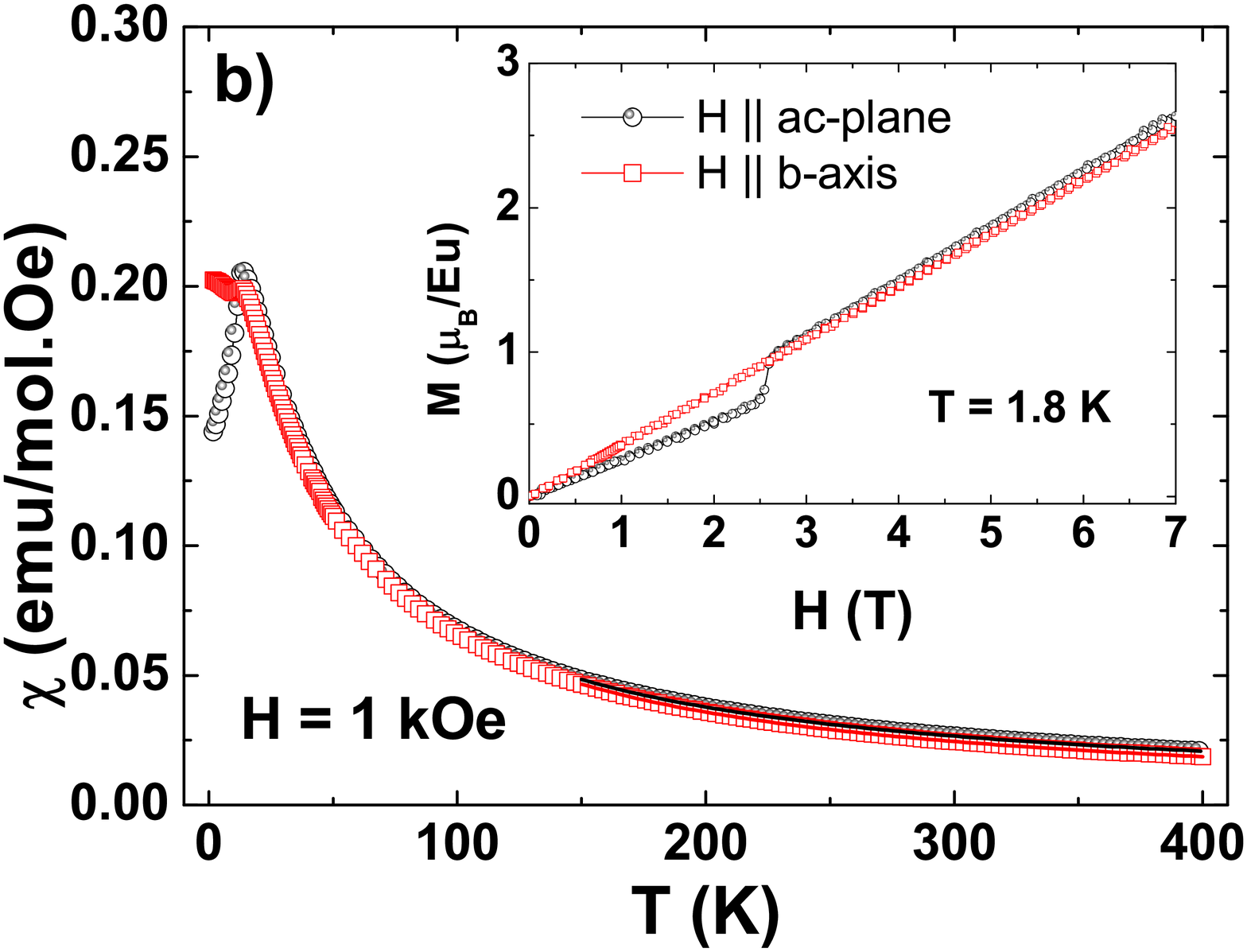}
\includegraphics[width=0.4\textwidth]{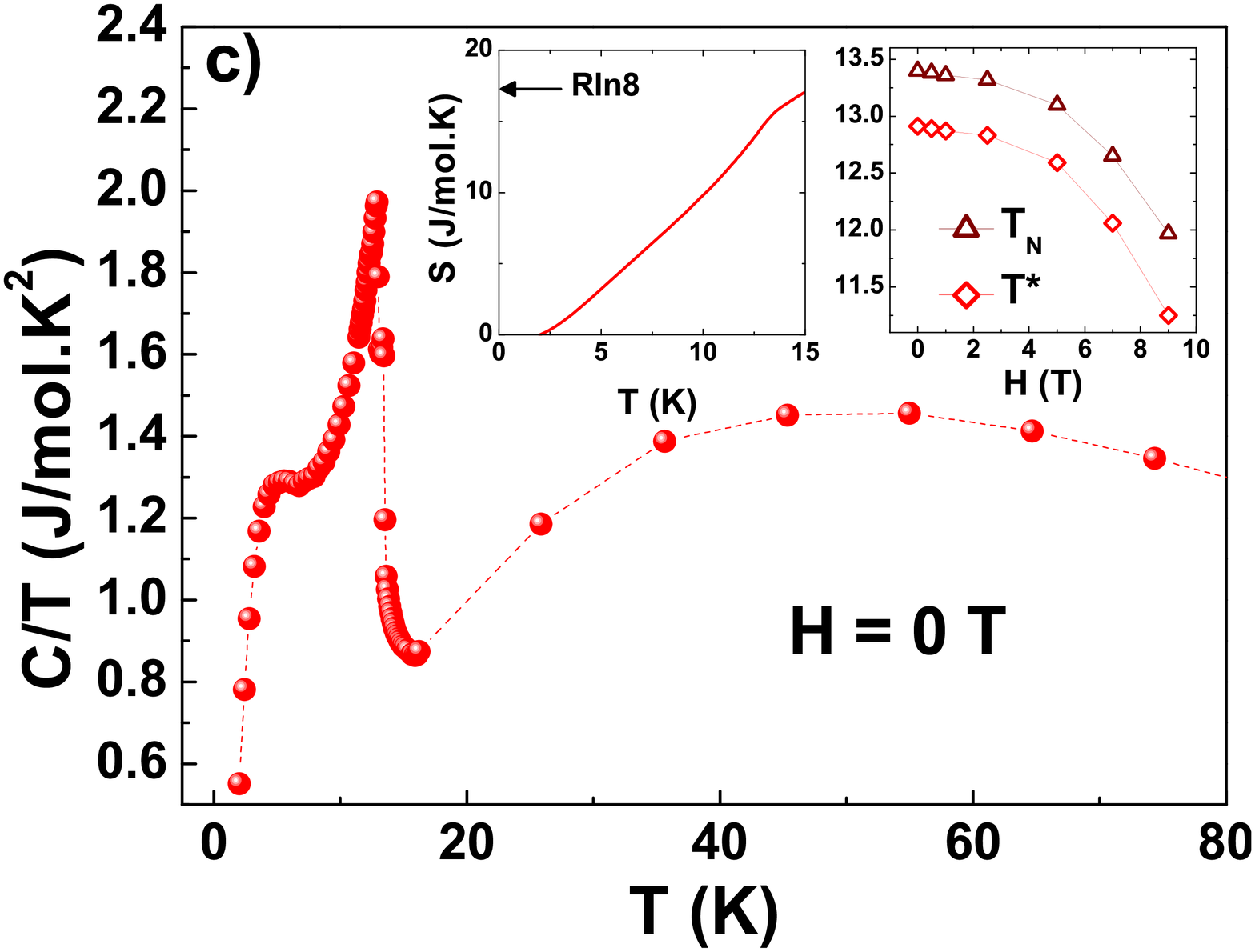}
\caption{Temperature dependence of macroscopic physical properties of EuPtIn$_{4}$ single crystals.
a) Electrical resistivity as a function of temperature for two current orientations. The inset shows the magnetoresistance for $H||$ $b$-axis at $T = 1.8$ K. b) Magnetic susceptibility with applied field H$=1$ kOe parallel to $ac$-plane and $b$-axis.
 c) Temperature dependence of specific heat. The insets show the recovered entropy (left) and the suppression of $T_{N}$ and $T^{*}$ with applied field along the $b$-axis.}
\label{Fig1}
\end{figure}

Fig. 2b shows the magnetic susceptibility as a
function of temperature for a magnetic field $H = 1$ kOe applied
parallel and perpendicular to the $ac$-plane of the sample.
$\chi(T)$ shows an isotropic
Curie-Weiss (CW) behavior at high-$T$ followed by an AFM
transition at $T_{N} = 13.3$ K. The sharp decrease of $\chi (T)$ below $T_{N}$ for $H ||$ ac-plane suggests that the $ac$ plane is the plane of easy magnetization.
From the CW magnetic susceptibility fits for $T>10T_{N}$
 (solid lines in Fig. 2b) we obtained for both directions a
CW temperature of $\theta_{CW}$ $\approx -15(1)$  K and an
effective moment of $\mu_{eff}\approx 7.8(1) \mu_{B}$ for Eu$^{2+}$ in
EuPtIn$_{4}$, which is in good agreement with the theoretical
value ($7.94 \mu_{B}$).  Isothermal magnetization curves as a function of the applied magnetic field at $1.8$ K are shown in the inset of Fig. 2b. On one hand, the magnetization increases linearly with field when $H ||$ b-axis, reaching an effetive moment of $2.5$ $\mu_{B}$ at $7$ T, yet below the full Eu$^{2+}$ moment of $7 \mu_{B}$. On the other hand, when $H ||$ ac-plane, a spin-flop transition is clearly observed at $H_{c} \sim 2.5$ T, also suggesting that the $ac$-plane is the plane of easy magnetization.

The AFM transition can also be observed in
lower panel c) of Fig. 2, which shows the specific heat per mole
divided by temperature. However, in this case it is possible to observe two close sharp peaks in \textit{C/T} at $T_{N} = 13.3$ K and $T^{*} = 12.8$ K. The former
corresponds to the onset of AFM order and its value is consistent with the
 magnetic susceptibility anomaly (see Fig.\ref{Fig1}b). The second peak at $T^{*} = 12.8$ K is likely related to a change of the magnetic structure. X-ray magnetic diffraction will help us to confirm this speculation. The estimated magnetic entropy recovered at
\textit{T$_{N}$} roughly reaches the value of $R$ln$8$ expected for
the whole Eu$^{2+}$ $S=7/2$ ion (left inset of Fig. 2c). Finally, both transitions are slightly shifted downwards with applied magnetic field (right inset of Fig. 2c).

Now we turn our attention to the microscopic properties of EuPtIn$_{4}$. In this regard, ESR is a highly
sensitive technique to study spin dynamics and magnetic
interactions and it often reveals details about the microscopic interaction J$_{fs}$ between the
4$f$ electrons and the \textit{ce} and about the Eu$^{2+}$--Eu$^{2+}$ magnetic correlations (see for instance references \cite{EuB6,EuC,EuRosa}).
Figure 3 shows the X-Band ($\nu = 9.4$ GHz) ESR spectra for both orientations measured at $T=100$ K. In both cases we observe a single ESR Dysonian resonance, consistent with a microwave skin-depth smaller than the sample size, indicating that the Eu$^{2+}$ ions in EuPtIn$_{4}$ experience	a	metallic	environment \cite{Dyson}. From the fitting of
the resonances to the appropriate admixture of absorption and
dispersion at $T=100$ K, we obtain a $g$-value of $g = 2.02(3)$ and linewidth
$\Delta H = 840(80)$ Oe for $H || ac$-plane and, for $H$ along the $b$-axis, $g = 1.97(4)$ and $\Delta H = 1100(100)$ Oe. Interestingly, even at high temperatures, the Eu$^{2+}$ ESR $\Delta H$ and resonance field $H_{res}$ are anisotropic, as shown in Fig. 3b, with smaller linewidths/resonance fields for $H||ac$-plane. Such identical variation is likely related to a g-value anisotropy due to crystalline electrical field (CEF) effects found for $S=7/2$ ions in orthorhombic systems \cite{Abragam}. Thus, the Eu$^{2+} $ESR linewidth results from an exchange narrowed $\delta$H with a distribution of local fields caused by unresolved fine (CEF splitting) and perhaps hyperfine structure. 
Using the $g$-value of Eu$^{2+}$ in insulators as 1.993(2) we extract an apparent small  $g$-shift which is negative ($\Delta$g $<$ 0) for H \textbar\textbar $b$ and positive ($\Delta$g $>$ 0) for H \textbar\textbar ac-plane \cite{Abragam}. This small effect may be an indicative of an anisotropic Eu$^{2+}$--Eu$^{2+}$ magnetic coupling with ferromagnetic
Eu$^{2+}$--Eu$^{2+}$ interaction in the $ac$-plane and antiferromagnetic Eu$^{2+}$--Eu$^{2+}$ interactions between the layers. However, we cannot rule out the contribution of demagnetization effects on this apparent $g$-anisotropy. Furthermore, there is a weak $T$-dependence of the g-values,
which may suggest the presence of short range magnetic correlations which yield non-trivial local fields at the Eu$^{2+}$ sites. However, we cannot rule out the contributions of `$bottleneck$" and ``$dynamic$" effects \cite{Rettori}. ESR experiments in the dilute series
Eu$_{1-x}$Sr$_{x}$PtIn$_{4}$ will help us to confirm this scenario and also to obtain further information about the relevant microscopic exchange interactions in this compound.

\begin{figure}[!ht]
\hspace*{0.5cm}
\includegraphics[width=0.5\textwidth]{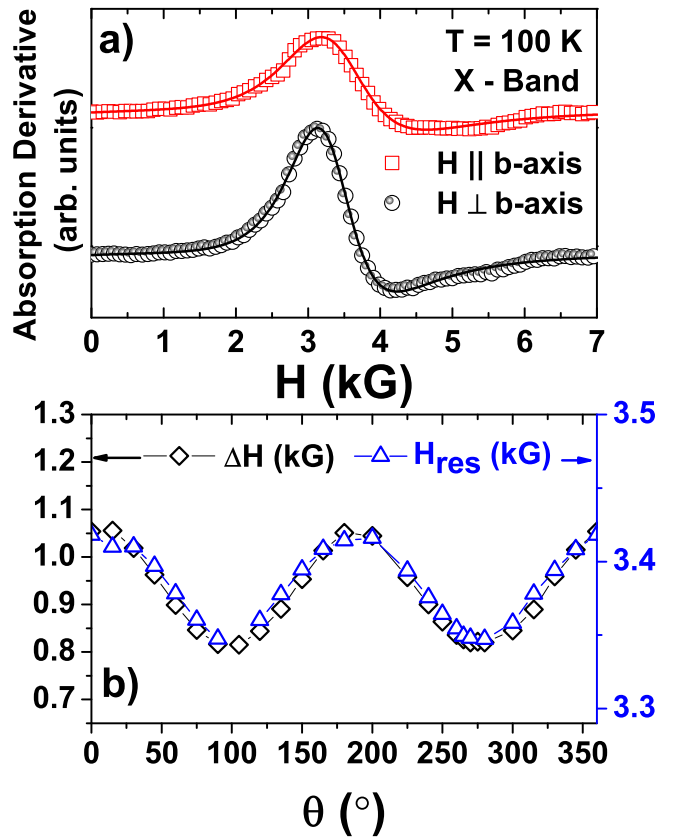}
\caption{a) X-Band ($\sim$ 9.5 GHz) ESR spectra of EuPtIn$_{4}$ single crystals at $T=100$ K. b) Angle dependence of ESR linewidth $\Delta H$ and resonance field $H_{res}$.}
\end{figure}

The $\Delta H$ and the $g$-value temperature dependence of the ESR
line of EuPtIn$_{4}$ for the X-Band is presented in
Figure 4a and 4b, respectively. An isotropic linear (Korringa) increase of $\Delta H$ with increasing-$T$ is observed for the Eu$^{2+}$ ESR signal in the paramagnetic state. From linear fits to $\Delta H (T)$ (solid lines) for $T > 100$ K, we extracted the value of the Korringa rate $b \equiv \Delta H/\Delta T = 4.1(2)$.
As the temperature is further decreased, the ESR $\Delta$H starts to broaden as
 a consequence of the development of short range magnetic correlations.  At the same temperature region, the g-factor slightly increases,  also suggesting the presence of a weak dominant ferromagnetic component.
Finally, below $T_{N}$ the resonance cannot be detected, likely due to the presence of antiferromagnetic collective modes which broadens the ESR line.

\begin{figure}[!ht]
\includegraphics[width=0.5\textwidth,keepaspectratio]{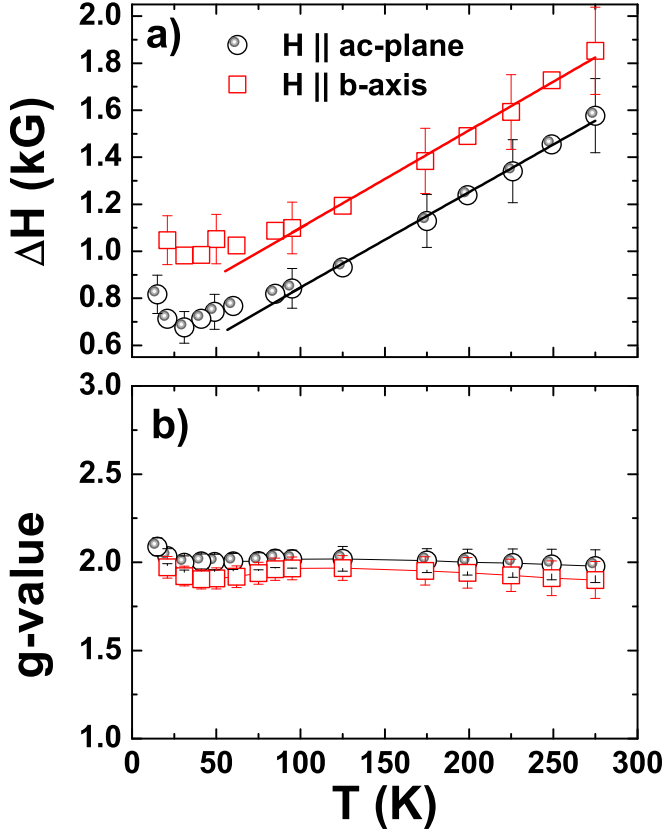}
\caption{Temperature dependence of Eu$^{2+}$ ESR $\Delta$H  and g-factor in X-Band.}
\end{figure}

To further explore the microscopic origin of the Eu$^{2+}$ ESR $\Delta$H anisotropy we have performed detailed electrical resistivity experiments in both paramagnetic and ordered regimes.  Fig. 5a shows the comparison between the anisotropy in Eu$^{2+}$ ESR $\Delta$H  and that in the electrical resistivity. Such comparison is important in this case because an anisotropic exchange interaction between the Eu$^{2+}$ 4$f$ electrons and the $ce$ would result in similar angular dependence of both physical quantities.
In fact, we observe that  both quantities display a similar anisotropy. This suggests the presence of anisotropic magnetic scattering due to anisotropic short range magnetic correlations between 
 Eu$^{2+}$ ions.
In addition, Fig. 5c shows a subtle change of anisotropy in the antiferromagnetic state as compared to the paramagnetic one (Fig. 5b), in agreement with the above picture. 
 
\begin{figure}[!ht]
\includegraphics[width=0.5\textwidth,keepaspectratio]{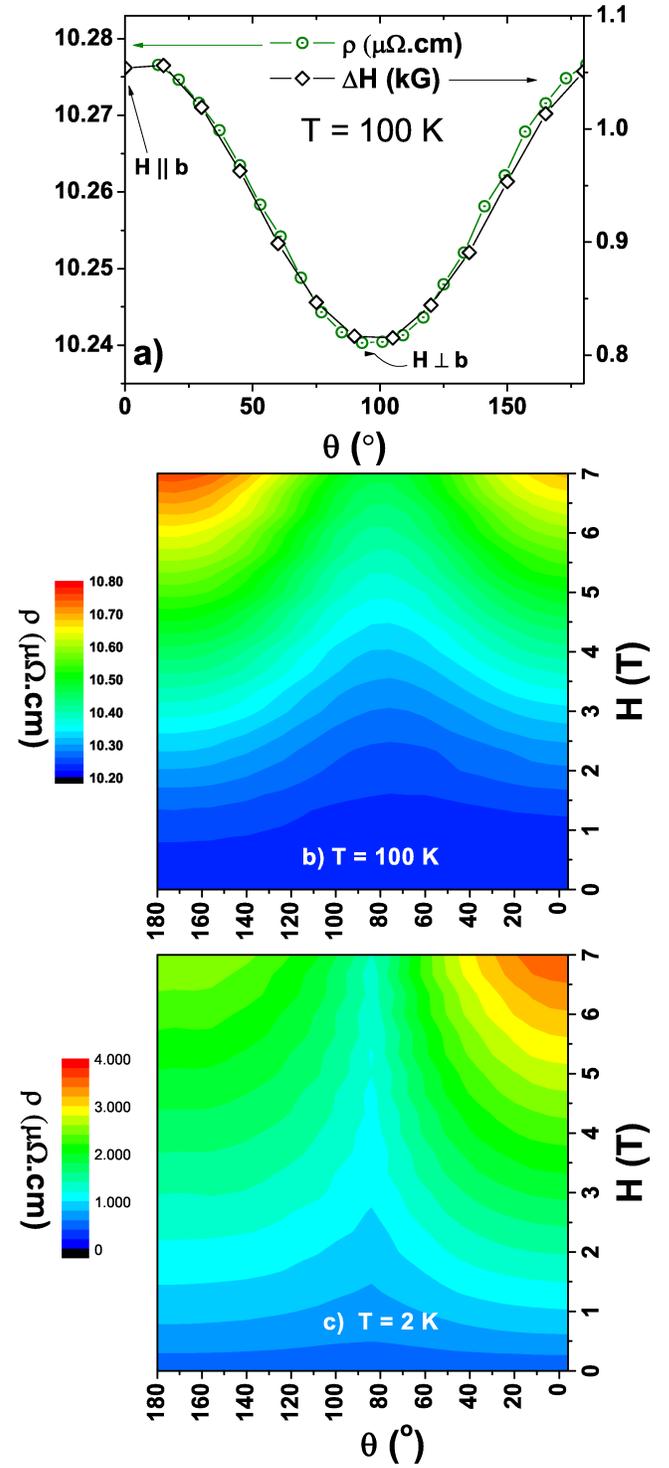}
\caption{a)  Angle dependence of in-plane resistivity and ESR linewidth. In-plane resistivity maps as a function of angle and magnetic field at b) $T=100$ K and
c) $T = 2$ K.}
\end{figure}

\section{Conclusion}

Here we report the synthesis, macroscopic characterization and ESR experiments on single
crystalline samples of EuPtIn$_{4}$. This compound crystallizes in an orthorhombic structure (space group Cmcm) and presents AFM ordering below T$_{N}=13.3$ K. A spin-flop transition is observed at $H_{c}\sim 2.5 T$ for magnetic fields applied along the $ac$-plane of easy magnetization.
In the paramagnetic state, a single Eu$^{2+}$ Dysonian ESR line with a Korringa-type relaxation is observed, indicating a metallic environment.
The anisotropy of ESR linewidth, resonance field and electrical resistivity at high-$T$ indicates the presence of both second order CEF effects and anisotropic exchange interaction between the Eu$^{2+}$ 4$f$ mediated by \textit{ce}. The latter may be caused by the low dimensionality of [PtIn$_{4}$] polyanionic networks surronding the Eu$^{2+}$ ions.

\begin{acknowledgments}
This work was supported by FAPESP, AFOSR MURI, CNPq, FINEP-Brazil.
\end{acknowledgments}

\end{document}